\documentclass{an}
\usepackage{graphicx}
\usepackage{times}
\usepackage{fancyhdr}
\usepackage{rotating}
\usepackage{hhline}
\usepackage{txfonts}
\begin{document}

\title{Weighing the cusp at the Galactic Centre}

  \author{N. Mouawad
          \inst{1}
          \and         
          A. Eckart\inst{1}
          \and
          S. Pfalzner\inst{1}
          \and
          R. Sch\"odel\inst{1}
          \and
          J. Moultaka\inst{1}
          \and
          R. Spurzem\inst{2} 
         }

   \institute{I. Physikalisches Institut, Universit\"at zu K\"oln,
               Z\"ulpicher Str.77, 50937 K\"oln, Germany
    \and              Astronomisches Recheninstitut, M\"onchhofstr. 12-14, 69120 Heidelberg, Germany
             }

\date{Received $<$date$>$; 
accepted $<$date$>$;
published online $<$date$>$}

\abstract{As stars close to the galactic centre have short orbital
periods it has been possible to trace large 
fractions of their orbits in the recent years.
Previously the data of the orbit of the star S2 have
been fitted with Keplerian orbits corresponding to a
massive black hole (MBH) 
with a mass of M$_{BH}$=3-4$\times$10$^6$M$_{\odot}$ 
implying an insignificant cusp mass.  
However, it
has also been shown that the central black hole resides in a $\sim$1''
diameter stellar cluster of a priori unknown mass.  In a
spherical potential which is neither Keplerian nor harmonic, orbits
will precess resulting in inclined rosetta shaped trajectories on the
sky.  In this case, the assumption of non-Keplerian orbits is a more
physical approach.  It is also the only approach through which cusp
mass information can be obtained via stellar dynamics of the cusp
members.  This paper presents the first exemplary modelling efforts in
this direction. 
Using positional and radial data of star S2, we find that there could exist
an unobserved extended mass component of 
several 10$^5$M$_{\odot}$ forming a so-called 'cusp'
centered on the black hole position. 
Considering only the fraction of the cusp mass M$_{S2_{apo}}$ within the 
apo-center of the S2 orbit we find as an upper limit that 
M$_{S2_{apo}}$/(M$_{BH}$ + M$_{S2_{apo}}$) $\le$ 0.05.
A large extended cusp mass, if present, is unlikely
to be composed of sub-solar mass constituents, but could be 
explained rather well by a cluster of high M/L stellar remnants, 
which we find to form a stable configuration.
   }

\correspondence{AN@aip.de}

\maketitle

\section{Introduction}
 Over the last decade, evidence has been found for the existence of
 massive black holes (MBH) in the centres of many nearby galaxies.
 With increasing observational data -- 12 MBHs candidates detected
 until 1995 (Kormendy \& Richstone 1995), and more than 37 until 2001
 (Kormendy 2001,  Ferrarese et al. 2001) -- it is argued that 
 most galaxies harbor nuclei dominated by MBHs with masses 
 that range between 10$^6$ and 10$^{9.5}$ solar masses.

 Located at a distance of only $\sim$8 kpc from the solar system (Reid
 1993; Eisenhauer 2003), the Galactic Centre (GC) is the closest and
 therefore best object for investigating physical processes in the
 galactic nucleus of a typical spiral galaxy.  It offers a unique
 ``laboratory'' for studying stars and gas in the sphere of influence on a
 super-massive black hole, (e.g., Genzel, Hollenbach, $\&$ Townes
 1994; Morris $\&$ Serabyn 1996; Mezger, Duschl, $\&$ Zylka 1996;
 Melia $\&$ Falcke 2001), with a degree of detail that cannot be
 accessed in any other galactic nucleus in the foreseeable future.
 With the now available high sensitivity and angular resolution, large
 ground-based telescopes offer the opportunity to obtain an
 unprecedented view of the Galactic centre.  Initially, with speckle
 imaging techniques and lately with adaptive optics techniques, high
 angular resolution images on the Galaxy's central cluster were
 obtained. This first set of observations was able to measure
 stellar motions on the plane of the sky, yielding estimates of the
 projected velocities (Eckart $\&$ Genzel 1996; Ghez et al. 1998),
 projected accelerations (Ghez et al. 2000; Eckart et al. 2002a), and
 three-dimensional orbital motions (Eckart et al. 2002a, Sch\"odel 
et al. 2002, 2003; Ghez
 et al. 2003b), which each provided a successively stronger case for a
 super-massive black hole at the centre of the Milky Way and its
 association with the unusual radio source Sgr~A* (Lo et al. 1985).

 More than a decade of high-resolution infrared observations of proper
motions in the GC, with the ESO New Technology Telescope (NTT) and the
ESO Very Large Telescope (VLT) (Eckart $\&$ Genzel 1996; Eckart et
al. \cite{eckart02}; Sch\"odel et al. 2002, 2003), as well as with the
Keck telescope (Ghez et al. 1998, 2000, 2003b), have revealed at least
6 stars that show substantial acceleration due to the super-massive
black hole associated with Sgr A* and are on bound orbits around it.
With a peri-centre of less than 0.6 mpc (15 mas) and an orbital period
of $\sim$ 15 years, S2 is the most striking case of these.

 A series of observations with the NAOS/CONICA adaptive optics
system/near-infrared camera at the ESO VLT unit telescope~4 that covered 
the peri-centre passage of the star S2 around Sgr~A* allowed 
Sch\"odel et al. (2002) to approximate a Keplerian orbit and to measure the 
enclosed dark mass down to a distance of $\sim$ 0.6 mpc from Sgr~A*.
 With these observations, they could exclude a
neutrino ball scenario (Munyaneza $\&$ Viollier, 2002) as an
alternative explanation for the dark mass concentration. They excluded as 
well a cluster of dark astrophysical objects (Maoz 1998) such as neutron
stars, leaving a central super-massive black hole as the most probable
explanation.

Using the Keck 10m telescope, Ghez et al. (2003a) 
measured a Keplerian orbit for the star S2 which agreed with the
results of Sch\"odel et al. (2002). Additionally, Ghez et al. (2003a) reported 
the first detection of spectral absorption lines 
(both Br$\gamma$ (2.1661 $\mu$m) and He I (2.1126 $\mu$m)),
providing the first line-of-sight velocity measurements of star S2.
 These measurements resolved the
ambiguity on the inclination of the S2 orbit indicating that its position was
behind the black hole when it passed through its peri-centre. In addition,
stellar rotational velocities suggest that S2 is an O8-B0 dwarf star
and thus a massive ($\sim$ 15~M$_{\odot}$) young star ($\le$ 10
Myrs).

 From data taken with NAOS/CONICA and the new NIR integral field
spectrometer SPIFFI at the ESO VLT, Eisenhauer et al. (2003) reported
new astrometric observations and additional spectroscopic observations
of the star S2, reducing the uncertainties on the orbital
parameters. They also gave the most accurate primary distance
measurement to the centre of the Milky Way of 7.94 $\pm$ 0.42 kpc,
which is in agreement with earlier determinations (see Reid 1993). 

In this paper, we explore the possibility that there exists a compact,
continuous mass distribution composed of several undetectable faint
stars or perhaps some more exotic material in addition to the point
mass of Sgr~A*. In this case, the orbit of S2 will not follow a simple
Keplerian orbit, but will rather show peri-centre-shifts that result in
rosetta shaped orbits.  This idea is motivated by the observation that
the stellar density does not flatten out, but exhibits a steep peak
towards the centre, a so-called cusp (Eckart et. al, 1995; Alexander
, 1999; Genzel et al., 2003).  In contrast to earlier studies,
the main approach of this work is that the mass-to-light ratio, $M/L$,
is {\it not} considered to be constant over the entire range of the GC
stellar cluster. 
Indeed, there are indications that the stellar 
population
varies with position and is not quite well mixed (Alexander, 1999).  
 Furthermore, the exact composition of the cusp is still unknown and 
our current
understanding of the stellar distribution in the GC is incomplete
(Genzel et al. 2003).
With current observations, low-mass stars 
($K$$\ge$ 21 mag) cannot be observed in these dense cusp regions and the
true value of the M/L ratio is not known.
 Also, as pointed out by Baumgardt et al. (2003), dynamical evolution of a
dense stellar cluster will result in a strong increase of 
M/L ratio by segregation of stellar evolution remnants to the centre.
 On the other hand, the approximation that the
dynamics in the central region is Keplerian is directly related to the
implicit assumption of Eckart et al. (2002a), Sch\"odel et al. (2002), and
Ghez et al. (2003a) that the $M/L$ ratio at 2~$\mu$m in the cusp is as
low as in the outer stellar cluster ($M/L = 2M_{\odot}/L_{\odot}$).\\
Considering these circumstances, stars with short orbital periods, in
particular the star S2, of which orbital data are best known, play a
key role in exploring the gravitational potential. We show that the
present observational data on S2 cannot discriminate between a
\emph{Keplerian} and a \emph{non-Keplerian} potential. Subsequently,
we study the influence of an extended distribution of dark mass near
Sgr~A* taking into consideration the constraints set by the orbit of
S2 as well as the limits set on the total enclosed mass 
at larger radii. 
In this study relativistic effects are neglected as they are
second order corrections.

In \textsection~2 we outline our modelling of the density distribution
of the central region and show that the star counts
near Sgr~A* can be described by a compact Plummer model core.
The method used to compute orbits in non-Keplerian potentials is 
explained in\textsection~3. In \textsection~4 possible
orbital models for S2 are presented. 
 We discuss the implications of the limits on the cusp mass 
derived  from the \emph{non-Keplerian} orbit modelling 
in \textsection~5, and draw our conclusions in \textsection~6.

\section{Modelling the stellar cluster and the enclosed mass}

In this section it is described how the most recent observations
of the GC star cluster (Genzel et al., 2000, 2003) can be used to model
the mass distribution function as close as a few milli-parsecs from
Sgr~A*.  A surface density profile is fitted to number density counts
and used, together with dynamical mass estimators, to
deduce the 3 dimensional mass distribution function.
By following this approach, we have assumed that the visible stars trace the
mass distribution of the cluster.

\subsection{Density model of the stellar cluster}
Genzel et al. (2003, see their Fig.7) determined the surface number
density of stars in the central parsec by counting the sources in
annuli with increasing radius around Sgr~A*. Within $\sim2$~arc-seconds
of Sgr~A*, their data indicate an excess of faint stars above the
counts expected in a cluster with a flat core, as it had been already
suggested earlier (Eckart et al. 1995; Alexander 1999). They confirmed
the existence of a "cusp" and showed that it is centred at a position
of ($\Delta$~$\alpha$, $\Delta$~$\delta$) = (+0.09'', - 0.15'') from
Sgr~A*, with an uncertainty of $\pm$ 0.2''.  Here we use the same
K-band counts ($K$ $\le$ 18) together with the earlier SHARP/NTT
counts ($K$ $\le$ 15), without taking into our study their higher
magnitude limit H-band counts ($H$ $\le$ 19).
In contrast to the 'by-eye' H-band counts, the K-band counts, were
determined by an automated procedure, they are therefore possibly less
biased, and cover the whole range of the cluster, not only the central
arc-second.

\begin{figure}
\centering
\resizebox{\hsize}{!}{\includegraphics[angle=270]{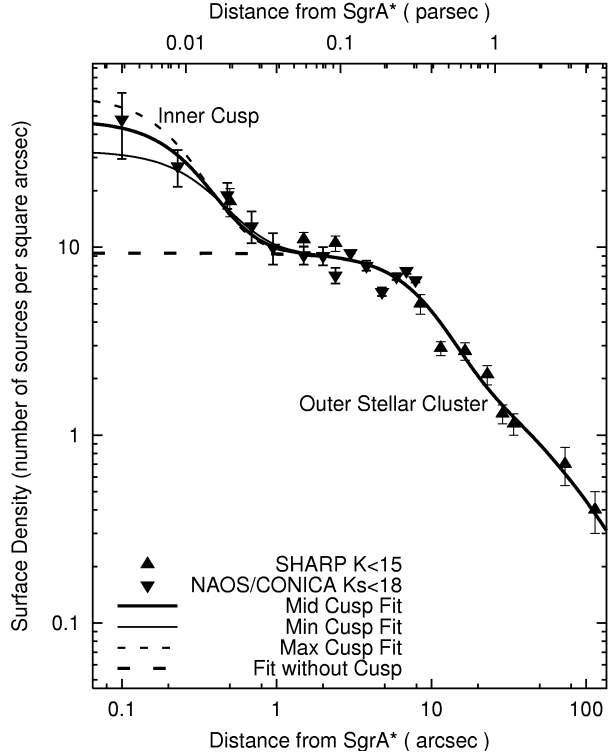}}
\caption{ Surface density of stars vs. distance from Sgr~A*. 
         The downward pointing triangles represent
        the CONICA/NAOS data for K$_s$$\leq$18, the upward pointing triangles
        are SHARP/NTT data for K$_s$$\leq$15 stars scaled up by a factor of 
        5 in order to match the fainter NAOS/CONICA counts.
        The data are taken from Genzel et al. (2000, 2003).
        The long-dashed line is the superposition of the 5 outermost Plummer
        models (see, Table~\ref{firstT}). The addition of the innermost 
        Plummer model (the 'inner' cusp') gives the 'composite model'. It
        is represented by the 3 other curves, where the thick-straight line 
        corresponds to the middle values of the 'inner cusp' parameters
        $R_1$ and $M_1$, the thin-straight line to its lower values, 
        and the short-dashed line to its upper values. 
        and the short-dashed line to its upper values. 
        }
\label{fig1}
\end{figure}

Genzel et al. (2003) fitted a broken power-law to their stellar
counts, with a distinct break at 10 arcsec. In this section, we show that
it is also possible to fit the number density counts with a
superposition of several Plummer models. 
The Plummer model has the advantage of being analytically integrable. 
It resembles actual clusters with compact cores and an extended outer 
envelope (Spitzer 1975).

In this model the 3 dimensional radial density distribution $\rho(r)$ 
and the projected mass density $\sigma(r)$ are described by:
\begin{eqnarray}
\rho(r)  & = & \frac{3M}{4 \pi R^3} \frac{1}{({1+{\frac{r^2}{R^2}})^{{\frac{5}{2}}}}} ,  \\
\sigma(r) & = & \frac{4 \rho(0) R}{3} \frac{1}{(1 +\frac{r^2}{R^2})^2} ,  \;  
\label{eq:2}
\end{eqnarray}
where $R$ is the core radius and $M$ the total mass. 

 Our best surface density fit to the data consists of a
superposition of 6 different Plummer models, which will be termed in the
following as 'composite model':

\begin{equation}
\sigma(r) = \sum^6_{i=1}{\frac{ M_i}{ \pi} \frac{R_i^2}{{(r^2 +
R_i^2)}^2}}.
\label{eq:3}
\end{equation}   

The six values of the core radii, $R_i$, are given in
Table~\ref{firstT}.  
Fits with larger numbers of Plummer models are possible as well;
six was the smallest numbers of components with which we were able to 
represent the data, i.e. the cusp plus the stellar cluster.

     \begin{table} 
     \caption[]{The parameters of the 'composite model' used to fit the
      GC stellar cluster. For each Plummer model we give the value of the core
      radius $R_i$ (in arcsec and in parsec) deduced from the surface
      density plot. $M_i$ is the corresponding total mass scaled to
      the enclosed mass distribution (Fig.~\ref{fig2}). 
      $M_{apo_i}$ is the contribution in mass inside a sphere which has a
      radius equal to the apo-astron of S2. $\rho_i$ is
      the calculated average density in M$_{\odot}$.pc$^{-3}$.
       The first Plummer model ($R_1$ =
      0.4'' or 0.015~pc) corresponds to the inner cusp centered at
      Sgr~A*. The addition of the remaining 5 models represents the
      density distribution of the outer stellar cluster.}
     \begin{tabular}[t]{||l||lllll||}
            \hline
     i &   $R_i$[arcsec] & $R_i$[pc] & $M_i$[M$_{\odot}$] & $M_{apo_i}$[M$_{\odot}$] & $\rho$$_i$(0)\\
            \hline
       1 &    0.4 &  0.015 &  8540        & 1080 & 6.0$\times$10$^8$    \\
       2 &    13  &  0.5   &  1.8$\times$10$^6$ & 10   &  3.5$\times$10$^6$     \\
       3 &    52  &  2.0   &  2.7$\times$10$^6$ & 0.25 &  7.9$\times$10$^4$      \\
       4 &    97  &  3.8   &  6.7$\times$10$^6$ & 0.01 & 2.9$\times$10$^4$    \\
       5 &    220 &  8.5   &  13.1$\times$10$^6$ & 0.002& 5.1$\times$10$^3$      \\
       6 &    321 &  12.4  &  27.6$\times$10$^6$&  0.001 & 3.4$\times$10$^3$      \\
         \hline
         \end{tabular}
         \label{firstT}
   \end{table}       

\subsection{The enclosed mass}

Similarly to eq.\ref{eq:3}, the mass distribution of the 'composite
model' is given by

\begin{equation}
M(r) =  \sum^6_{i=1}{\frac{M_i r^3}{(r^2 + R_i^2)^{3/2}}}.
\label{eq:4}
\end{equation}

Under the assumption that the visible stars trace the mass,
the M$_i$ can be determined by fitting eq.\ref{eq:4}
to the dynamical enclosed mass estimates.

In order to calibrate the enclosed mass M(r)
we only rely on a solid measure
of the central mass (within 0.001~pc from SgrA*)
derived from the Keplerian orbit of the star S2
and mass estimates of the entire stellar cluster
at distances of several parsecs from the center.
At these large radii the published results of all different statistical
mass estimators give very
consistent results within uncertainties that are
fully acceptable for our study
(Eisenhauer 2003, Genzel et al. 2000, Serabyn $\&$ Lacy 1985; 
G\"usten et al. 1987; Lacy et al. 1991; Herbst et al. 1993; Roberts $\&$ Goss 
1993). By tying our mass modelling as a function
of radius to the stellar number density distribution, 
it does not depend on possible anisotropies of the velocity field
at radii of less than a parsec.
It is only there where the anisotropies are apparently present at a 
considerable level
(Genzel et al. 2000, Sch\"odel et al. 2003).
In particular we would like to point out that we are not
fitting to enclosed mass estimates at these small radii.
In Fig.2 these are only shown for completeness.
The only assumption that goes into our considerations
is that of a constant M/L ratio as a function of distance.
This assumption is relaxed and discussed
in the following sections of the paper.

Our best fit to these dynamical mass estimates corresponds 
to a 2.9$\times$10$^6$~M$_{\odot}$ $\pm$
0.3$\times$10$^6$~M$_{\odot}$ black hole plus the mass distribution of
eq.\ref{eq:4}, of which the 6 total masses $M_i$, are listed together
with their corresponding densities, $\rho_i$, in Table~\ref{firstT}.
 An offset in the value of the enclosed
compact mass will not influence our modelling significantly because
the shape and mass of our cluster composite model depends primarily on
the profile of the stellar number counts.

The combination of the five outermost components
gives a similar configuration as the flattened isothermal sphere of
core radius $\sim$ 0.3 pc derived by Genzel et al. (2003) (see mass densities
below).
Fig.~\ref{fig1} shows that there exists a mini-core centered on Sgr~A*, 
which we denominate by the 'inner cusp'. It is represented by the innermost
 Plummer model with a core radius $R_1$ = 15.5mpc $\cong$ 0.4''.

With an apo-centre of approximately 9~mpc (Sch\"odel, 2004; Ghez et al., 
2003a), the star S2 orbits in a region
inside the core radius ($R_1$$\sim$15mpc) of this innermost
model. Therefore its path can be mainly influenced, in addition to the BH 
potential, by the 'inner cusp' gravitational potential 
(see colomn 5 in Table~\ref{firstT}).
At these distances from the centre, the present number density counts
have large error bars (see Fig.~\ref{fig1}).  The limits on the core
radius, $R_1$, of our model, deduced from the error bars on the number
density counts, are found to be 13.2~mpc for the lower value 
$R^{min}_{1}$, and 20.2~mpc for the upper one $R^{max}_{1}$. 
Varying the mass, $M_1$, of the central Plummer model component 
allows to study cases in which the mass-to-light ratio, $M/L$, 
 varies as a function of distance from Sgr~A*.          
As described in \textsection~ 3 and \textsection~ 4, exploring the possible
ranges for $R_1$ and $M_1$ that fulfill the constraints given by the 
orbital measurements of star S2 allows to derive dynamical 
information on the cusp mass.
In the following, we make use of this model which fixes the density profile
of the cusp to the stellar profile.

\begin{figure}
\centering
\resizebox{\hsize}{!}{\includegraphics[angle=270]{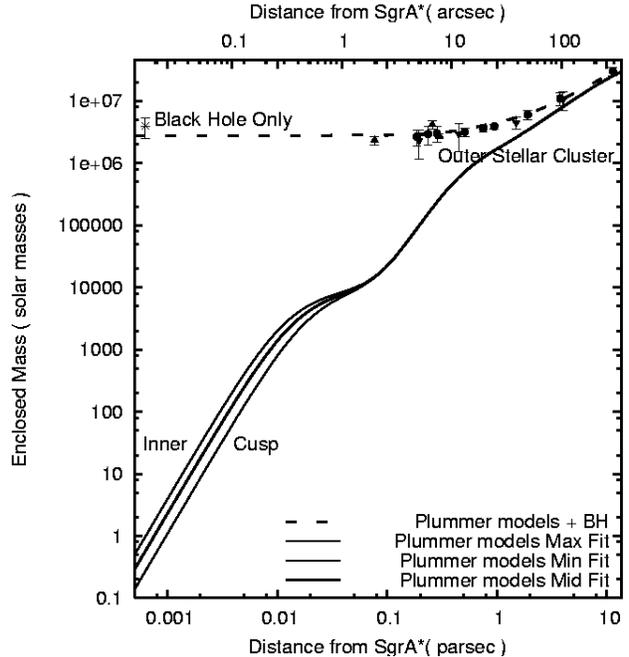}}
\caption{Mass distribution
in the central 10 pc of the Galactic Centre for an 8 kpc distance
(Reid et al. 2003; Eisenhauer et al., 2003).The
star point denotes the mass derived from the orbit of S2. The error
bar combines the orbital fit and astrometry errors (Eisenhauer et
al. 2003). Filled circles are mass
estimates from a parameterised Jeans-equation model of the radial
velocities of late type stars, assuming isotropy (Genzel et al. 2000,
Sch\"odel et al. 2002).
We omitted all mass estimators at radii of less then 1'' since they 
are effected by the surface mass density of the cusp which had not been taken
into account in previous investigations.
Filled upward-pointing triangles denote Leonard-Merritt projected
mass estimators from a new NTT proper motion data set (Sch\"odel et al. 2002). 
Filled down-pointing triangles denote mass estimates from the ionized
 and neutral gas dynamics (Serabyn $\&$ Lacy 1985; G\"usten et al. 1987; 
Lacy et al. 1991; Herbst et al. 1993; Roberts $\&$ Goss 1993).
A comparaison of representative mass models to the measured velocity 
dispersions is given in Genzel et al. (1996, 2000).
  The long-dashed line is a fit to the measured enclosed
mass against distance. That results from a combination of the mass
distributions (see \textsection~2.2) deduced from the composite model 
fit to the
stellar cluster, plus a 2.9$\times$10$^{6}$ M$_{\odot}$ black hole.
The same model without a point mass is represented by the 3 other
curves, they correspond to acceptable models 
of the stellar cluster within the error bars.}
        
\label{fig2}
\end{figure}

In Fig.~\ref{fig2} the long-dashed line corresponds to our best fit,
the thick-straight line shows only the enclosed mass without a BH
obtained with our model of the extended cluster.  
The thin-straight and the short-dashed lines
represent the above discussed extreme cases of an inner cusp radius
$R_1$ = $R^{min}_1$ = 13.2~mpc with $M$ = 8540$_{\odot}$ and $R_1$ = 
$R^{max}_1$ = 20.2~mpc with $M$ = 8820$_{\odot}$. Note that these errors 
result from the uncertainties of the stellar number counts at the smallest
distances to Sgr~A* (Fig.~\ref{fig1}).

The density of the 'inner cusp' is
a few hundred times larger than that of the second Plummer model and
it has a $\sim$ 33 times smaller core radius (see Fig.~\ref{fig1} and
Table~\ref{firstT}).
Our model implies density values as high as
1.68$\times$10$^8$ M$_{\odot}$pc$^{-3}$ at 0.1", and
1.54$\times$10$^7$ M$_{\odot}$pc$^{-3}$ at 1''. These values are
comparable to the ones derived by Genzel at al. (2003) from a broken
power-law density profile (7$\times$10$^8$
M$_{\odot}$pc$^{-3}$ at 0.1"and 3$\times$10$^7$ M$_{\odot}$pc$^{-3}$
at 1'').

In the following we will use the above derived mass distribution to
determine the path of S2 in response to this non-point mass potential
plus the central black hole.

 \section{Modelling the orbit of S2}

In order to fit non-Keplerian orbits to the measured time-dependent
positions of the star S2 in the above described type of mass
potential, one needs a high accuracy integrator. For this purpose, we
chose a fourth order Hermite integrator derived from the one used in
high-accuracy $N$-body simulations (Aarseth 1999, Spurzem 1999, for
an introduction of the Hermite scheme see Makino $\&$ Aarseth
1992). The advantage of the Hermite scheme is that it allows a fourth
order accurate integration based on only two time steps. Therefore it
requires the analytic computation of the time derivative of the
gravitational force. This is the point where the integrability of the
Plummer model is convenient. We computed the trajectory of S2 around
the BH and through the extended mass as given by our mass model (see
\textsection~2). The presence of an extended mass induces retrograde
pericentre-shifts which result in open rosetta shaped orbits.  

In total there are ten fit parameters - six orbital elements, the
total central mass M$_{tot}$, the fraction $f$ of the extended mass to
the total mass M$_{tot}$, and the 2-D position $P$ of the central
mass. All these values are known within a certain margin.  We used the
position measurements for S2, the dynamical position of Sgr~A* and
the 5 line-of-sight velocity measurements as determined by Sch\"odel et 
al. (2003), Eisenhauer et al. (2003) and Ghez et al. (2003a).

A valid computed orbit has to fit both the measured velocities and
positions. Therefore, two reduced $\tilde{\chi^2}$ values, one for the
positions, $\tilde{\chi^2}$$_{pos}$, and another for the velocities,
$\tilde{\chi^2}$$_{vel}$, were estimated. While fitting the orbits, 
we considered the $\tilde{\chi^2}$$_{pos}$
values for the positions and velocities independently because the
$\tilde{\chi^2}$ values for the velocities were, in general, relatively 
much smaller than for the positions. 
In some cases they indicated that measurement errors were systematically
overestimated. 
This way, it was possible to attain, both in velocities and in 
positions, a similar fit quality.
Under this condition, it is always possible to define a common 
${\chi^2}_{tot}$ by adding both the $\chi^2$ values obtained from the 
positions and the velocities.
The total reduced $\tilde{\chi^2}$, $\tilde{\chi^2_{tot}}$, is simply the 
${\chi^2}_{tot}$ divided by the number of degrees of freedom which is 
equal to 33 in our case.
Table~\ref{secondT} lists obtained total reduced $\tilde{\chi^2_{tot}}$ and 
non-reduced $\chi^2$$_{tot}$ which indicated errors $\le~5\sigma$.
These errors were estimated by applying a deviation $\Delta$$\chi^2$$_{tot}$
from the lowest obtained $\chi^2$$_{tot}$.
In our following analysis we used orbital fits with an error $\le$~1$\sigma$.

   \begin{table} 
       \caption[]{Characteristic $\tilde{\chi^2}$$_{tot}$ of our fits. 
       The first column
       lists the obtained characteristic $\tilde{\chi^2}_{tot}$ for 
       which the resulting orbits are considered to be fitting orbits.
       The 2$^{nd}$ column expresses the total non-reduced $\chi^2$$_{tot}$
      for 33 degrees of freedom 
      ($\tilde{\chi^2}$$_{tot}$ = $\chi^2$$_{tot}$/33).
      When column 3 lists the deviation of the obtained $\chi^2_{tot}$ 
      from the lowest obtained $\chi^2_{tot}$, column 4 corresponds their 
      resulting $\sigma$ error bars.}
       \begin{tabular}[t]{||l||l||l||l||}
             \hline
          $\tilde{\chi^2}$$_{tot}$    & $\chi^2$$_{tot}$ & $\Delta$$\chi^2$$_{tot}$  & $\sigma$\\
         \hline
          0.68  &  22.5 & 0  & $<$ 1$\sigma$\\
          0.71  &  23.5 & 1  & 1$\sigma$\\    
          0.80  &  26.5 & 4  & 2~$\sigma$\\
          0.95  &  31.5 & 9  & 3~$\sigma$ \\
          1.17  &  38.5 & 16 & 4~$\sigma$ \\
          1.44  &  47.5 & 25 & 5~$\sigma$\\
             \hline
          \end{tabular}
          \label{secondT}
    \end{table}

\subsection{Influence of model parameters}

In order to study the main effects of a varying cusp mass on the
orbit of the star S2 we investigated a representative 
coverage of the parameter space that includes the full 
range of possible scenarios. In each single orbital run, the total central
mass, M$_{tot}$ and its position as well as the extended mass fraction, $f$, 
were fixed; only the 6 orbital elements are varied. 
The three parameters, M$_{tot}$, the central mass position and $f$,
were varied independently forming a fully representative 3-D grid.
In total $\sim$360 orbits were computed and fitted to the data. In the 
following, we give a short description of this parameter space:\\

The total mass M$_{tot}$, together with the the three velocities and
three coordinates determine the \textbf{orbital elements}.
From the imaging data, sky positions and proper motions are given in a
certain error range for each epoch.  
For a chosen initial epoch, these six  
input values are varied within their observational uncertainties. 
Our minimization method is a grid scheme, which, although expensive 
in computing time (time $\propto$ N$^6$, N being the number of steps per 
parameter), is a robust method which does not fall easily on local minima.
Choosing an appropriate $N$ for each parameter,  $\chi${$^2$} values were 
computed and a 6-D grid was constructed. Ultimately, the lowest $\chi${$^2$} 
value determines the best fitting orbit.

The black hole is treated as a point mass with the extended component being centered on it. The lower central mass limit 
(Reid et al. 2003) and upper central mass limit (Ghez et al. 2003; Sch\"odel et al. 2003) are well determined. Based on these earlier works, 
the \textbf{value of the total central
 mass} is varied between 2.6~$\times$10$^{6}$M$_{\odot}$ 
and 4.8~$\times$10$^{6}$M$_{\odot}$, considering six different values 
of $M$$_{tot}$.
In addition, for any of these given \textbf{total central masses values,
$M_{tot}$}, the effect of various \textbf{extended mass fractions, $f$}, 
was investigated (see Fig.~\ref{fig4}).

\begin{figure}
\centering
\resizebox{\hsize}{!}{\includegraphics[angle=0]{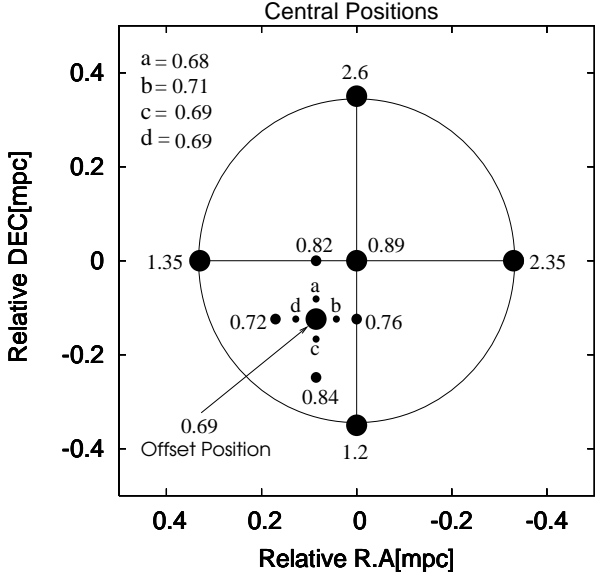}}
\caption{The different positions of the central mass as they were considered
associated with the lowest obtained $\tilde{\chi}$$^2$$_{tot}$ value at each 
given position. The \emph{offset} position is centred in the area showing the 
lowest $\tilde{\chi}$$^2$$_{tot}$ values.}
\label{fig3}
\end{figure}

We used the position of Sgr~A* as determined by Reid et al. (2003)
with an uncertainty of 10~mas.  
The \textbf{location of the central mass} was varied at 14 different positions within 
this error bar. Fig.~\ref{fig3} shows these positions along with their
obtained $\tilde{\chi}$$^2$$_{tot}$ values. These were fairly 
distributed around the \emph{offset} position which is the deviation of the 
focus of the Keplerian orbit from the central radio position the way 
determined by Eisenhauer et al. (2003).
In order to check
the accuracy of the focus of the Keplerian orbit, we compare the two
positions for the star S2 at the epoch 2002.30 calculated with the
orbital parameters given by Eisenhauer et al. (2003) and by Ghez et
al. (2003b). From the corresponding
positions and uncertainties, an offset of the projected orbital focus given by 
Ghez et al. (2003) from the one by Eisenhauer et al. (2003) of
$0.0\pm2.9$~mas East and $0.8\pm2.0$~mas North can be derived.  
Hence, we find an excellent agreement between the two independent results.

  In order to take into account the uncertainty of the \textbf{core radius
 of the inner cusp}, the orbital fits computed for our best fitting model
value of $R_1=15.5$~mpc, were repeated only in the case of the fitting orbits
with an error $\le~2~\sigma$, 
for the extreme cases of $R^{min}_1=13.2$~mpc and $R^{max}_1=20.2$~mpc.

\section{Results of non-Keplerian fits}

\subsection{Orbits with lowest ${\chi}${$^2$}{$_{tot}$} values}
Our results show that the available data on the path of S2 can be fitted 
very well with \textbf{\em non-Keplerian orbits}. However, at present the orbit 
of S2 is not sufficiently well determined to exclude a pure Keplerian orbit
(see Fig.~\ref{fig6}).  In fact, in both cases a minimum  
$\tilde{\chi}${$^2$}{$_{tot}$} value of $\sim$0.7 was obtained. 
In other
words, it cannot be excluded that there exists an extended mass 
component in addition to the black hole.
In the case of Keplerian orbits and for different central mass position, fits for a 3.65$\times$10$^6$M$_{\odot}$ black hole shows
 the lowest $\tilde{\chi}${$^2$}{$_{tot}$} value.
If we assume an additional extended mass component, 
 the lowest $\tilde{\chi}${$^2$}{$_{tot}$ values corresponded to 
fits for a total mass of 3.65$\times$10$^6$M$_{\odot}$ of which 5\% are 
extended or a 4.1$\times$10$^6$M$_{\odot}$ total mass of which 10\% or 15\% 
are extended, or 4.45$\times$10$^6$M$_{\odot}$ total mass of which 20\% are extended, 
as well as a 4.8$\times$10$^6$M$_{\odot}$ total mass of which 25\% are extended.
We can also consider only the fraction of the cusp mass M$_{S2_{apo}}$ 
which is located at distances from the MBH smaller than the 
apocenter of the S2 orbit. In this case the 25\% limit corresponds to 
$$M_{S2_{apo}}/(M_{BH} + M_{S2_{apo}}) \le 0.05. $$
}
The orbital parameters for these $\le$1$\sigma$ fitting orbits
 are given in Appendix~A together with the full parameter
range that produces orbits with an error of $\le$5$\sigma$ . Within the
uncertainties the orbital
parameters for the Keplerian case agree with earlier work (Sch\"odel et
al. (2003), Ghez et al. (2003), Eisenhauer et al. (2003)).

\begin{figure}
   \centering
\resizebox{\hsize}{!}{\includegraphics[angle=0, height=6cm]{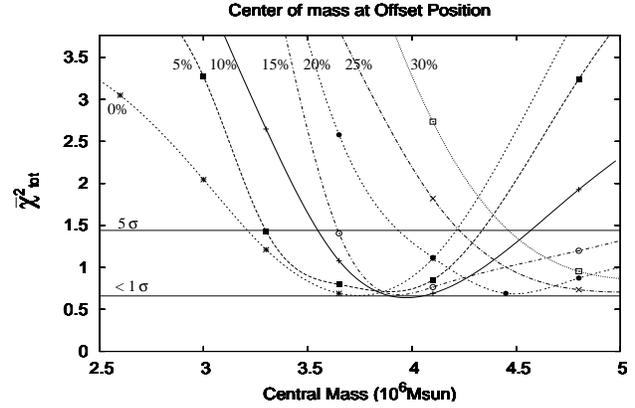}}
\caption{Plot of the $\tilde\chi^2_{tot}$ values vs. the total central
mass, M$_{tot}$, calculated for different extended mass fractions,
$f$, with the point mass at the \emph{offset} position.  Values of the same
$f$ are fitted with splines.  Crosses: $f$ = 0; ``x'' symbols: $f$ = 0.05;
stars: $f$ = 0.1; empty squares: $f$ = 0.15; filled squares: 0.2,
empty circles: $f$ = 0.25; and filled circles: $f$ =
0.3. 
The two horizontal lines denoted by \emph{$<$1~$\sigma$} and \emph{5~$\sigma$} 
delimit the region of fitting orbits ($\tilde\chi^2_{tot}$ $\leq 1.44$).}
\label{fig4}
\end{figure}

\subsubsection{Position of Sgr~A*}

 \begin{figure}
\centering
\resizebox{\hsize}{!}{\includegraphics[angle=0]{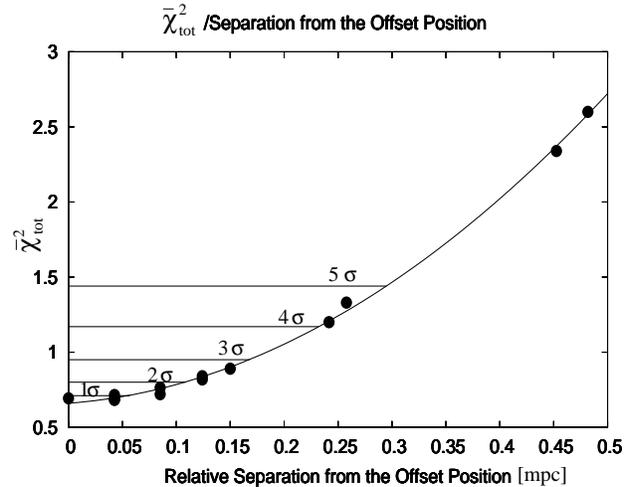}}
\caption{The lowest obtained reduced $\tilde{\chi}$$^2$$_{tot}$ values in function of the separation from the \emph{offset} position. These are fitted with a function of the form 7.2x$^2$ + 0.5x + 0.7.}
\label{fig5}
\end{figure}

Only orbits with focuses at $a$, $b$, $c$, $d$ and \emph{offset} positions 
(see Fig.~\ref{fig3} and Fig.~\ref{fig5})
resulted with an error of $\le$1$\sigma$. This gives a constraint on the 
SMBH black hole position of only a separation of $\pm$0.05~mpc ($\pm$1.3~mas) 
in R.A or in Dec. from the \emph{offset} positions.

   \begin{figure*}
   \centering
   \includegraphics[width=17cm]{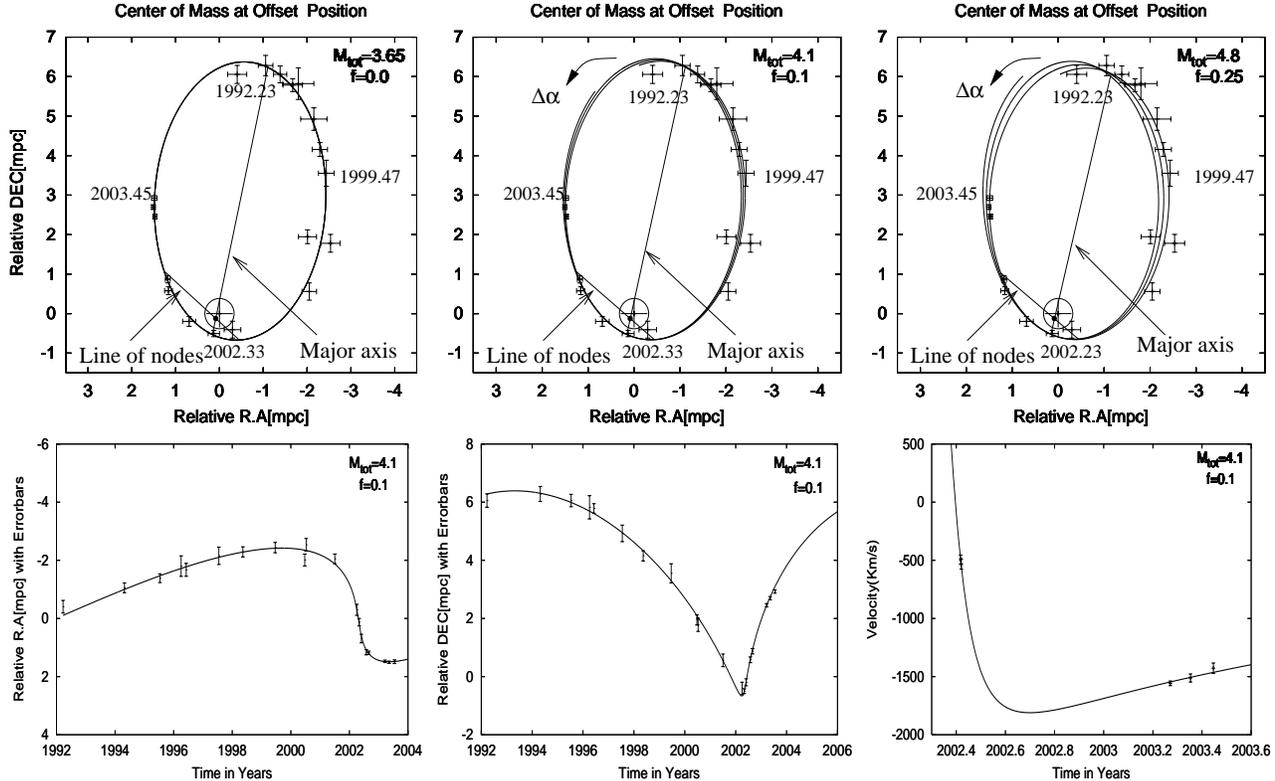}
   \caption{Three exemplary orbits determined by our fitting. Upper
left panel: Keplerian orbit with 3.65$\times$10$^6$M$_{\odot}$ point
mass. Upper middle panel, non-Keplerian orbits with
4.1$\times$10$^6$M$_{\odot}$ total mass, thereof 10\% extended.  Upper
right panel, non-Keplerian orbits with 4.8$\times$10$^6$M$_{\odot}$
total mass, thereof 25\% extended.  Here the central mass is at the
offset position, 0.09mpc east and 0.12mpc south from the nominal
radio position of Sgr~A*.  The lower panels show the velocity in the
relative R.A, the relative Dec., and along the line-of-sight as a
function of time for the case of 4.1$\times$10$^6$M$_{\odot}$ total
central mass with a 10$\%$ extended component.  This case is
representative for the other cases. The direction of $\Delta$ $\alpha$
the pericentre-shift is shown by an arrow.}
    \label{fig6}
    \end{figure*}

\subsubsection{Higher total central masses and cusp masses}

Fig.~\ref{fig4} shows a plot of $\tilde{\chi^2}_{tot}$ vs.\ different discrete
central total mass values and for each fraction $f$ of the extended
mass at the offset position.  We also examined orbits with total
masses higher than 4.8$\times$10$^6$~M$_{\odot}$, up to
11.6$\times$10$^6$~M$_{\odot}$.  
 We found that at a 5~$\sigma$ level of confidence
 the S2 orbit could fit the data for
cusp masses of up to 70\% of the total mass. 
This resulted in an upper
 limit of cusp masses as high as 8$\times$10$^6$~M$_{\odot}$ 
corresponding to an extended component of $\sim$1$\times$10$^6$~M$_{\odot}$
inside the sphere of a radius equal to the apocentre of S2.  However, the
total masses in these cases are in disagreement with the dynamical
mass estimates at large radii. Therefore, we regard them as 
unrealistic but mention the result for completeness.  

Large amounts of extended mass inside
the S2 orbit would be possible if the mass distribution function at
the centre were steeper than in our model or if the core radius were
smaller than the apo-centre of S2 ($\sim9$~mpc).  In that case 
we found that mass densities -- exceeding 
 the currently derived few 10$^8$~M$_{\odot}$pc$^{-3}$ 
by three or four orders of magnitude -- are unlikely to form a 
stable configuration because stellar collisions
will become of increasing importance (see section 5.2).

\subsubsection{Core radius and slope of the density profile}

In the case of the \emph{offset} position, for all the orbits with 
errors $\le$~2~$\sigma$,
the fitting
procedure was repeated for the extreme cases of $R^{min}_{1}$=13.2~mpc and
$R^{max}_{1}$=20.2~mpc of the inner cusp radius.  Even with these
different core radii, the case of the \emph{offset} position 
is found to give the lowest $\tilde\chi^2$$_{tot}$ values. 
Also the fit quality remains unchanged at $\tilde\chi^2$$_{tot}$ $\simeq$~0.7.
We also consider that the compact component 
can have a much smaller core radius than the 
stellar profile, possibly well within the S2 apocentre. In this case, following
the result of Rubilar \& Eckart (2001, see their Fig.5), we deduce that 
for the same extended mass fraction, $f$, the induced pericentre-shift 
is smaller for a 
bigger core radius. Thus for a core radius that does not exceed 
a value 15.5~mpc, our estimation on the mass of the cusp 
is an upper limit.\\

Using a series of Plummer model density profiles,
implies a value of the cusp slope of $\alpha$ $\le$~5.
As mentioned in \textsection~2,
Genzel et al. (2003) fitted the stellar number density with a broken power-law
of a slope equal 1.4$\pm$1.
From theoretical models of cusps around massive black holes, one would 
expect power-law indecies in the range 3/2  $\le$  7/4
(Bahcall \& Wolf 1977; Murphy et al. 1991).
To study the influence of the denstiy profile on the pericentre-shift,
we performed fittings of the S2 orbit considering a power-law profile of a 
slope equal to 1.4.
The pericentre-shift induced by the same amount of mass inside S2's apocentre
but different fractions of mass, $f$,  
did not exceed the one induced by a Plummer model density profile 
by more than 12\%. Thus, by covering
 a wide range of slopes 1.4 $\le$$\alpha$$\le$ 5, 
we found that for the shallower mass distribution,  
the pericentre-shift never exceeds 12\% of our initially derived value, thus 
our estimates indicate an upper limit on the mass that could be
present inside the S2 apocentre, and consequently on the mass of the cusp.\\

\section{Discussion}

In the previous sections it was demonstrated that the measured positions and
line-of-sight velocities of S2 are not sufficiently well 
determined to exclude a Keplerian orbit. Assuming a simple Keplerian case, the central
gravitational potential by definition is entirely dominated by the 
point mass associated with Sgr~A* and no dynamical constraints can be 
derived on any extended mass component due to e.g.\ the surrounding 
stellar cluster. 
More revealing, however, is the more physical assumption of 
non-Keplerian orbits. 
The presence of a stellar cusp shows that -- at least to a certain extent
-- there is some extended mass present near the black hole.
This potential can be modelled by a central black hole 
plus an extended mass component. We assumed that the 
cusp has the same shape as the observed stellar 
number density counts which we fitted by a Plummer density model 
(see \textsection{2}).
We also took into account that the total enclosed mass
at larger radii (1pc) would not exceed a value of 
$\sim$~4.8~$\times$~10$^6$~M$_{\odot}$ (see section~3.1).

Having shown that the current data allow for 
an extended mass component higher than the one of 8540~M$_{\odot}$ 
inferred from stellar number density counts which were scaled to match
 the velocity distribution at large radii (see \textsection{2}),
 we will discuss its nature
in sections \textsection{5.1} and \textsection{5.2}.
 The value of 8540~M$_{\odot}$, 
 representing the total mass of the 'inner cusp', is
equivalent to a mass of 3100~M$_{\odot}$ inside the core radius $R_1$
(15.5~mpc or 0.4''). It is based on the assumption that the ratio of
the stellar number counts to total stellar mass does not vary with
radius and environment in the Galactic Centre, i.e. that the
mass-to-light ratio $M/L$(2$\mu$m) is constant (see also Genzel et al.,
2003). It is very likely that this assumption is unjustified. Genzel et
al. (2003) find that the stellar population in the cusp differs to a
certain degree from the population of the surrounding, large-scale
cluster. Also, effects such as mass segregation and stellar collisions
might work very effectively in the dense environment of the
cusp and $M/L$ would show higher values towards the centre 
(Baumgardt et al. 2003).
Therefore, we consider that the mass-to-light ratio, $M/L$
(2$\mu$m) of the 'inner cusp', can be different from that of the outer 
stellar cluster.
In the following, two hypotheses will be discussed: The existence of a
cluster of faint low-mass stars, not yet detectable with the
resolution and sensitivity of current instruments, and the existence
of a cluster of heavy dark objects like stellar black holes and
neutron stars.

\subsection{Extending the K-band Luminosity Function to Faint Stars}

We use the $K$-band luminosity function (KLF) for K$\leq$ 18 mag
within a projected radius of $1.5''$ from Sgr~A* as it was determined
by Genzel et al. (2003 - see their Fig.11). 
Note that the KLF is not corrected for extinction or for the distance 
modulus. 
The authors fitted the KLF with a power-law with a slope of
$\beta$=0.21 $\pm$ 0.03, ($\beta$ =dlogN/dK). Since this result is
based on a number of roughly 60~stars within $1.5''$ of Sgr~A*, the
observed stars themselves cannot account for a significant extended mass
component. It is therefore necessary to extrapolate the KLF at the
faint end. However, we assume that the observed stars trace the mass
carrying population which can currently only be accessed 
theoretically by extrapolating the KLF towards its faint end.

For the extrapolation of the KLF two different slopes are considered,
$\beta$ = 0.21 and a steeper one, $\beta$ = 0.35, which fits the Bulge
KLF (Alexander \& Sternberg, 1999) as well as a model of an old
($\sim10$~Gyr) stellar cluster of solar metalicity 
(Zoccali et al. 2003). In the following we concentrate on the
region within R$_1$ of the inner Plummer model component.
The total number of stars in the inner cusp (R$\leq 0.4''$ or 15.5~mpc) 
brighter than a given magnitude can be directly
estimated from the (extrapolated) KLF.

The KLF gives the number of stars per surface area per magnitude. 
In order to calculate the total number of stars and the total 
stellar mass present within the spherical volume enclosed by the 
core radius of R$ = 0.4''$  of our central Plummer model component,
we de-projected the KLF.
Table~\ref{clusters} lists the resulting
numbers of stars, the total cluster mass, M$_{Cl}$, the average
stellar mass in the cluster, M$_{aver}$, and the corresponding
$M/L$ (2$\mu$m) for clusters given by KLFs with slopes of $\beta=0.21$ and
$\beta=0.35$ and for different cut-off magnitudes, between K$= 20$
(M$_{min}$ = 1M$_{\odot}$) and $K$ = 28 (M$_{min}$ = 0.06M$_{\odot}$).
M$_{Cl}$ is calculated from the observed KLF using A$_K$=3 and the 
$M/L$~(2$\mu$m) values listed in Table~\ref{stars}.

The numbers listed in Table~\ref{clusters} show that the mass of stars
present within $R$ = 0.4'' deduced from a $\beta=0.21$ slope KLF cannot 
be higher than 800 M$_{\odot}$, even after extrapolation to the faintest
magnitudes (see Table~\ref{stars} for a list of K-magnitudes and0.1051
corresponding MS-stars at the GC).  Therefore, a cluster with a KLF of
$\beta=0.21$ cannot explain a mass of 3100 M$_{\odot}$ within 
R$ = 0.4''$, estimated from direct number density counts with an 
$M/L$(2$\mu$ m) = 2~M$_{\odot}$/L$_{\odot}$ (see \textsection~2).

$M/L$~(2$\mu$m) converges to a value 2~M$_{\odot}$/L$_{\odot}$
for $\beta$ = 0.35, and a
magnitude limit between 26 and 27, in this case we find the value of 3100
M$_{\odot}$ required by our mass distribution modelling in \textsection{2}.
 Lower magnitude limits would
increase $M/L$~(2$\mu$m) further, and the mass of the inner cusp calculated
from our Plummer model would be underestimated.  Stellar types
corresponding to the required M/L~(2$\mu$m) ratios are listed in
Table~\ref{stars}.

Table~\ref{clusters} also shows that for a main-sequence stellar
population M/L(2$\mu$m) does not exceed
$4.0$~M$_{\odot}$/L$_{\odot}$, even if we consider very
faint/low-mass stars ( $K$ = 28) and a steep faint-end KLF 
($\beta$ = 0.35). 
This value corresponds to a cusp representing only 0.2$\%$ of the total mass
(see Table~\ref{T5}).
We conlude that the amount of extended mass allowed
inside the core radius of the 'inner cusp' would be far too large to 
be explained by a stellar population of MS
stars, and therefore would require another type of mass carriers.  
We show in \textsection~5.3 that such a configuration is possible.

\begin{table}
\caption{Extinction free mass-to-light ratios, $M/L$(2$\mu$m), for
        different stellar types in luminosity class V at the distance
        of the Galactic Centre.}
\begin{tabular}{||l||l||l||l||l||}
\hline
M/L(2$\mu$m)[M$_{\odot}$/L$_{\odot}$] & K[mag] & Mass[M$_{\odot}$] & Spectral types \\
\hline
1.29   & 16.425 & 1.60 & F0 \\
1.93   & 17.005 & 1.40 & F5 \\
2.73   & 17.695 & 1.05 & G0 \\
5.02   & 18.665 & 0.79 & K0 \\
5.36   & 18.915 & 0.67 & K5 \\
11.77  & 20.065 & 0.51 & M0 \\
9.23   & 20.065 & 0.4  & M2 \\
16.66  & 20.915 & 0.33 & M3 \\
38.49  & 22.315 & 0.21 & M5 \\
87.55  & 23.815 & 0.12 & M7 \\
166.43 & 25.265 & 0.06 & M8 \\  
\hline
\end{tabular}
\label{stars}
\end{table}

\begin{table*}
\caption{For each 'observed' K-magnitude limit (not corrected for
extinction) we give the numbers of stars $N$, cluster mass M$_{Cl}$,
average stellar masses M$_{aver}$, and mass-to-luminosity ratio,
$M/L$(2$\mu$m), within the central $0.4''$ for hypothetical Plummer
model type clusters with KLFs extrapolated by two power-law slopes of
$\beta=0.21$ and $\beta=0.35$.  M$_{Cl}$ is the mass inside the core
radius R$_{1}$ of the innermost Plummer model component. The derived
$M/L$ values include an extinction correction for A$_K$=3.}
\begin{tabular}{||l||llll||llll||}
\hline
& & & $\beta$= 0.21 & & & &  $\beta$=0.35 &  \\ 
\hline
Magnitude limit& N & M$_{Cl}$[M$_{\odot}$] & M$_{aver}$[M$_{\odot}$] & 
M/L(2$\mu$m)[M$_{\odot}$/L$_{\odot}$] &
 N & M$_{Cl}$[M$_{\odot}$] & M$_{aver}$[M$_{\odot}$] & M/L(2$\mu$m)[M$_{\odot}$/L$_{\odot}$]\\
K $\leq$ 28 & 3500 &800 & 0.9 & 1.0 & 44500 & 5200 & 0.1 & 3.7 \\
K $\leq$ 27 & 2200 & 710& 1.4 & 0.9 & 20000 & 3700 & 0.15 & 2.8 \\
K $\leq$ 26 & 1350 & 620& 2.3 & 0.8 & 9000 & 2400 & 0.3 & 1.8 \\
K $\leq$ 25 & 820  & 525& 3.7 & 0.7& 4100 & 1500 & 0.8 & 1.1  \\
K $\leq$ 24 & 510  & 450& 6.0 & 0.6& 1920 & 1000  & 1.6 & 0.8 \\
K $\leq$ 23 & 310  & 390& 10  & 0.5& 950 & 700 & 3.3 & 0.5  \\
K $\leq$ 22 & 190  & 325& 16  & 0.4& 510 & 480 & 6.0 & 0.4 \\
K $\leq$ 21 & 120  & 235& 26  & 0.4& 320 & 380 & 9.7 & 0.3 \\
K $\leq$ 20 & 70   & 193 & 42  & 0.3& 230 & 270 & 11.2 & 0.2 \\

\hline
\end{tabular} 
\label{clusters}
\end{table*}   

\begin{table}
\caption{Mass-to-light ratios $M/L$(2$\mu$m), for the different
\emph{offset} position fitting orbits obtained from our model
calculations.  The errors on M/L are model dependent. From scaling to
the enclosed mass estimate in Fig.~\ref{fig2} we estimate an
error of 5\%. Columns 3, 4, and 5 give the cusp mass inside the 
orbit of S2 for core radii of respectively 20.2mpc, 15.5mpc and 13.2mpc.} 
 \begin{tabular}{||@{}l@{}||l||l||l||l||@{}l@{}||}
\hline
M$_{tot}$         &  $f$ & M$_{S2_{apo}}$&M$_{S2_{apo}}$& M$_{S2_{apo}}$&  M/L(2$\mu$m)\\
$[10^6 M_{\odot}]$&      & R$_1$=        & R$_1$=       & R$_1$=        & [M$_{\odot}$/L$_{\odot}$] \\
 & & 20.2mpc & 15.5mpc  & 13.2mpc &  \\
\hline
3.65 & 0.0 & 0.001 & 0.001 & 0.001  & 2.00\\ 
3.65 & 0.05 & 0.011 & 0.021 & 0.031 & 42.74   \\
3.65 & 0.10 & 0.022 & 0.044 & 0.062 & 85.48  \\
4.10 & 0.05 & 0.013 & 0.024 & 0.033 & 48.01 \\
4.10 & 0.10 & 0.025 & 0.049 & 0.071 & 96.02  \\
4.10 & 0.15 & 0.038 & 0.071 & 0.104 & 144.03 \\
4.10 & 0.20 & 0.052 & 0.098 & 0.140 & 192.04  \\
4.45 & 0.20 & 0.049 & 0.105 & 0.155 & 227.7 \\
4.80 & 0.20 & 0.061 & 0.108 & 0.157 & 224.82 \\
4.80 & 0.25 & 0.069 & 0.135 & 0.190  & 269.32  \\
4.80 & 0.30 & 0.089 & 0.164 & 0.237  & 337.24 \\
\hline
\noalign{\smallskip}
\end{tabular}
\label{T5}
\end{table}

\subsection{Stability of a Cluster of Low-Mass Stars}

For a multi-mass stellar distribution, high mass stellar remnants
(stellar black holes and/or neutron stars) are expected to migrate to
the centre as a consequence of dynamical friction.  One would expect
that, within a Hubble time, these compact objects show a higher
concentration toward the centre than the lighter ones (Morris et al.,
1993; Miralda-Escud$\acute e$ $\&$ Gould, 2000), which should be transferred
by that mechanisms to orbits at greater distances from the centre of
the cluster. This argues against the existence of
a cluster of low-mass stars in the inner cusp.

On the other hand, the possibility of a cluster of low-mass stars
 cannot be excluded and we are far
from understanding the properties of the stars in the cusp. There is
for example the unexplained presence of massive young stars,
e.g. MS O/B-type stars close to the
black hole (Genzel et al., 1997; Eckart, Ott \& Genzel, 1999; Figer et
al., 2000; Gezari et al., 2002; Ghez et al., 2003).  
These stars have not had enough time to achieve energy equipartition
with the fainter older stellar population. They are hence
dynamically un-relaxed. Also, there are
indications for a radial anisotropy of the stars in the cusp which
might be un-relaxed (Sch\"odel et al., 2003), in
spite of the expected short relaxation time in this dense environment.
Because of this general lack of theoretical understanding of the
cluster near Sgr~A*, we consider that a cluster of faint/low mass stars
should not be ruled out entirely from the possible interpretations of
the inner cusp.

\subsection{Is the Cusp Dominated by Dark and Massive Objects?}

In this section, we consider orbital fits with errors of 
$\le$1$\sigma$ that result in high cusp masses yet are in good agreement 
with the enclosed mass measurements.
The evaluation presented in \textsection~5.1 shows that such a heavy cusp is 
unlikely to consist of stars only.
Here we study whether such a cusp could consist of heavier 
mass carriers like stellar black holes or neutron stars.

\subsubsection{Presence of stellar remnants in the center 
due to dynamical friction}
When the compact remnants of massive stars are themselves 
significantly more
massive than the normal field stars in the Galactic Centre, as would be the 
case of black hole remnants, than they are susceptible to inward migration
as a consequence of dynamical friction. The resulting mass segregation can lead
to a pronounced concentration of compact objects within the central stellar
core within a Hubble time (Morris, 1993).
$N$-Body simulations of globular clusters showed that the combination 
of stellar evolution (production of stellar mass
 black holes, neutron stars, white dwarfs, and of binaries including
 such objects) and stellar dynamics will almost certainly lead to a
 strong increase of $M/L$ in the central parts of the nuclear star
 cluster. Black holes and neutron stars sink to the centre and
 may coalesce (G\"urkan et al. 2003).
Such highly detailed self-consistent simulations of the dynamical 
episodes are, however, not yet possible on the scale of the Galaxy.
 $N$-Body simulations (even with cutting-edge special purposes computers like
GRAPE-6 (Makino 2001))
 cannot follow the evolution of a galactic nucleus over a Hubble time if 
relaxation is appreciable (Freitag \& Benz, 2002).

Under certain assumptions and for some range of initial parameters, 
less realistic but more efficient 
Monte Carlo numerical simulations of the evolution of the GC 
(Freitag \& Benz, 2002) showed that the SBHs sink to the center on a short timescale 
of a few gigayears, settle into a centrally concentrated distribution and 
dominate the stellar mass there.
Similarly, Murphy Cohn $\&$ Durisen (1991, see their figure 8b) showed that 
densities higher than 10$^9$ M$_{\odot}$~pc$^{-3}$ could reside in regions
as close as few mpc from the central SBH.
 Other studies by Morris (1993) and
Miralda-Escud\'e $\&$ Gould (2000) estimate that 10$^4$ - 10$^5$ SBHs, due to
dynamical friction, would have accumulated at distances of less than 
about 1 pc from the center.
These latter works can only account for about 10$^2$ - 10$^4~$~M$_{\odot}$ 
within separations of about 20 mpc from the centre which represents only 
1$\%$ to 10$\%$ of our upper mass limit derived from the orbit of S2.
However, the central density of the SBH cluster depends on various uncertain 
quantities: the SBH mass function, the stellar IMF and formation rate, 
the remnant progenitor masses, and the dynamical age of the GC.
Morris (1993) argued that within a wide range of assumptions about the IMF, 
and about the minimum stellar mass capable of producing a black hole remnant,
the total mass of remnants concentrated into the inner few tenth of a parsec,
would be about 0.4-5 10$^6$ M$_{\odot}$. 
 If the black hole remnants were to achieve equipartition with the field stars,
they would form an inner core with a radius of a size as small as 50~mpc.
These would coalesce or form a quasi-stable cluster of stellar mass black
 holes.
Even if a catastrophic merger of stellar remnants did occur at some point at
the age of the galactic nucleus, the continuous influx of massive 
stellar remnants would ensure that a concentrated population of them is present
within the stellar core.
These findings apply, however, in the absence of an initial central 
black hole.
To our knowledge, a similar study of a cluster with already a pre-existing 
super-massive black hole, as in the centre of our Galaxy, is not yet 
available.

\subsubsection{In situ formation of stellar remnants}
In the case of the Galactic Centre, complex dynamical episodes
have taken place. A numerous variety of young early-type, bright and massive 
stars exist 
at distances of 10~mpc - 400~mpc from the centre. There exist a dozen of 
bright O/B main sequence stars within about 40 mpc of Sgr~A*, these are
fast moving S-stars (mass $\sim$ 20~M$_{\odot}$) of which the star S2 is an 
example. There exists also $\sim$~30 more massive 
(30~M$_{\odot}$~$\lesssim$~mass~$\lesssim$~100~M$_{\odot}$) very bright 
early-type stars, the so-called He stars exhibiting He/HI emission lines.
The existance of these two types of stars at these regions from the center
is still not understood. These could have formed there or would have migrated 
there due to different in-spiraling processes (Genzel et al. 2003; Ghez et 
al. 2003; Gerhard et al. 2001, Krabbe et al. 1995).
If such stars were always present in the center of the Galaxy, due to stellar
evolution, their remnants would contribute very efficiently to the
formation of a dense cluster also at the center.
Here, we assume that these stars would end their lives in the 
formation of neutron stars or stellar black holes with masses between 
1.5~M$_{\odot}$ and 10~M$_{\odot}$ (on average 7~M$_{\odot}$), 
we also consider their lifetimes to be $\lesssim$~10$^7$ yrs.
If after a single lifetime about 40 of such stars form stellar remnants, 
throughout the age of the Galaxy ($\sim$ 10$^{10}$ yrs), it is
possible to account for about 10$^5$~M$_{\odot}$ needed to explain our
upper mass limit of the 'inner cusp'.

Considering the two above described scenarios - while these do not 
represent the complete history of the galactic nucleus - it is, however,
 a fair conclusion that \textit{strong} 
to \textit{moderate} dynamically caused $M/L$ variations prevail at the GC.

Our attempts in this work is not to explain the formation
of a high density cluster of stellar remnants at the GC. 
We would rather like to analyse such a configuration if existant.
In the following we investigate if such a hypothetically high $M/L$ 
configuration of stellar remnants can form a stable configuration.\\

\subsubsection{Stability of a cluster of stellar remnants}
Rauch $\&$ Tremaine (1996) studied the configuration of a central massive 
black hole plus an extended mass distribution $M$ of radius $R$ consisting 
of objects with mass $m$ in terms of its $non-resonant$ relaxation time 
$t_{rel}^{nr}$.
Under the assumption that $M$ $<<$ $M_{BH}$, Rauch $\&$ Tremaine (1996) 
derive how $t_{rel}^{nr}$ depends on $M_{BH}$, $M$, and the orbital 
time scale $t_{orb}$ at the outer edge of the cluster.
If 
(a) the stellar orbits have random orientations and moderate eccentricities,
(b) the density of stars is approximately uniform within $R$, and 
(c) $M_{cluster}$ $/$ $M_{BH}$ = 10$^{-2}$ - 10$^{-5}$ 
then  Rauch $\&$ Tremaine (1996) find that

\begin{equation}
t^{nr}_{rel} \sim \frac{M_{BH}^2}{m^2N ln\Lambda} t_{orb},
\label{eq:6}
\end{equation}               

 where $ln \Lambda$ is the Coulomb logarithm $\sim$~13 in this case.
This situation should - to first order - be applicable to the Black
Hole/cusp scenario at the Galactic centre.  Condition (a) is probably
fulfilled with the possible exception that the stars in the cusp might
have fairly high eccentricities (Sch\"odel et al. 2003).  Assuming a
Plummer model as a cusp description fulfills condition (b). 
We consider the case
where a peri-centre-shift of the order of 1$^{\circ}$ is induced.
As an example, we choose the case of a cusp mass of 15\% with a total
mass of 4.1~$\times~10^6$~M$_{\odot}$. 
This cusp mass is well below the upper limit derived in section 4.1. 
In the case of a cusp mass of 15\%, $M/M_{BH}$ is of
the order of 10$^{-2}$, if we restrict ourselves to the region within
the core radius $R_{1}$.  This is close to what is required by
condition (c).
 
Here, $M$$_{BH}$ = 3.5 $\times$ 10$^6$ and $m_{sr}$ is the average mass 
of the stellar remnants.
A value of $m_{sr}$ = 5 M$_{\odot}$ is roughly consistent with a composite 
cluster made of neutron stars ($m$ $\sim$ 1.5~M$_{\odot}$) 
and stellar black holes (5~M$_{\odot}$ $<$ m $<$ 25~M$_{\odot}$).
We then find that $t_{rel}^{nr}$ $\sim$ 10$^6$ $\times$ t$_{orb}$,
if we assume that most of the mass inside the core radius 
is present in the form of stellar remnants of average mass $m_{sr}$.
Considering that the core radius of  15-20~mpc will define t$_{orb}$ 
 we find $t_{rel}^{nr}$ to be about $2\times 10^7$ years.
\\
\\
We can also investigate whether such a configuration is stable by 
estimating how many stellar black holes evaporate.
Integrating a Maxwell distribution function for the velocities
above the escape velocities gives the
percentage of stars not bound to the BH.
For the velocitiy dispersion, following Alexander et al. (2003),
we can write:
\begin{equation}
v_{escape}=  \sqrt{2}v_{circular} = \sqrt{2(1 + \alpha)}~\sigma .
\label{eq:7}
\end{equation}              

For a value of $\alpha$ = 2, about 0.03\% of the present stellar black
holes will be evaporated after each relaxation time.  For a steeper
cusp with $\alpha$=3 we find that 0.006\% of the stellar black holes
will evaporate on that time scale.  For $t_{rel}^{nr}$ of a few 10$^7$
years and $\alpha$ between 2 and 3 about 50\% of the stellar black
holes will have evaporated after 25 to 250 $t_{rel}^{nr}$
corresponding to about $5\times10^8$ to $5\times10^{9}$ years.  Given
the fact that there will also be an influx of mass i.e. stellar black
holes or neutron stars from outside the cusp one can consider such a
configuration stable over a significant fraction of the Milky Way's
age.

Given equation (\ref{eq:6}), for a fixed $t_{orb}$,
the relaxation time decreases linearly with $N$, i.e with the cusp mass.
Thus, cusp masses well exceeding 2~$\times~10^5$~M$_{\odot}$ would 
not form a stable configuration compared to the age of the Milky Way.

\section{Summary}
From stellar number counts, we have a strong reason to assume that there
is some extended mass around Sgr~A*. Therefore, orbits of stars near
Sgr~A* should be analyzed in terms of non-Keplerian motion. Only in
that way it is possible to find out more about the composition of the
cusp. This paper presents the first approach to this problem.  In this
work we studied the case where the contribution of the central stellar
cluster at the centre of the Milky Way could introduce
peri-centre-shifts of the order of 1 degree per revolution on the orbit
of S2.  We assume that the dynamics of S2 is governed by both the
super-massive black hole and the central cluster.  The steeply rising,
cusp-like stellar density distribution in the inner few tens of
milli-parsecs near Sgr~A* can be modeled with a Plummer model
distribution having a relatively small core radius of the order of the
S2 semi-major axis. In a spherical potential, the stellar orbits will
precess and the stars will follow rosetta shaped trajectories (Binney
$\&$ Tremaine, 1987; Rubilar $\&$ Eckart, 2001; Fragile, \& Mathews,
2000).  We performed orbital fits to the S2 data for this class of
orbits. We used a fourth order Hermite integrator, adequate for the
required precision.  The runs were performed for a large range of
initial parameters for the total mass, the position of the centre of
mass, the fraction of the extended mass component, and the 6 classical
orbital parameters.

We found that the S2 trajectory can be equally well fitted by 
$non-Keplerian$ orbits as by $Keplerian$ orbits. For the Keplerian case, we
could confirm the results of Sch\"odel et al. (2003), Ghez et
al. (2003), and Eisenhauer et al. (2003).  Basically independent of the 
amount of total or extended mass we find consistently a value for the 
central Black Hole mass of about $3.7\times10^6$M$_{\odot}$.
This is well between but not coincident with one of the two possible
Black Hole mass estimates of either $4.75\times10^6$M$_{\odot}$
or $2.79\times10^6$M$_{\odot}$ that were derived by Aschenbach et al. (2003)
from an analysis of the quasi-periodicity observed in strong 
X-ray flares.

A $non-Keplerian$ orbit with a fit error of $\le$~1~$\sigma$ that shows a peri-centre-shift  of the order of 1$^{\circ}$ implies a total mass of 
4.1$\times$10$^6$ M$_{\odot}$ with a fraction of the extended mass 
being $f=0.15$.
With this value of $f$, the amount of
mass present in the inner cusp is $\sim$75 times larger than the one
expected form a direct mass estimation deduced from the stellar
density and the integrated mass plot ( see Fig.~\ref{fig1} and
Fig.~\ref{fig2}) while assuming a constant 
2$\mu$m mass-to-light ratio of 2~M$_{\odot}/$L$_{\odot}$. 

If the inner cusp is mainly composed of main sequence stars,
then a $M/L$~(2$\mu$m) of 2~M$_{\odot}/$L$_{\odot}$ can only be attained in
the case of a faint-end KLF slope of $\beta$ = 0.35 and a magnitude
limit 25 $<$ $K$ $<$ 26. The mass of such a cluster will be 
3100 M$_{\odot}$ within $R$ = 15 mpc derived from our
initial Plummer model in \textsection~2.  The existence of even
fainter ($K$ $<$ 28), low-mass stars (0.06 M$_{\odot}$) in the inner
cusp will not contribute more than $\sim$3 times the mass in the
inner cusp; such a configuration is not likely to be stable for a long time.
In addition it would be unclear how these stars would get into the
inner cusp. It is unlikely that they formed there, however, they might have
migrated into that region by loosing angular momentum by interactions with
stars in the overall Galactic Centre stellar cluster with a 
core radius of 0.3-0.4pc.

The example of a total mass of 4.1$\times$10$^6$M$_{\odot}$
and an extended mass component of 15\%, as discussed above, 
implies an $M/L$($2\mu$m) 
of $\sim$ 145 $\pm$ 10~M$_{\odot}/$L$_{\odot}$ ; 
here objects with large $M/L$ values
are needed to explain the extended cusp mass.  One plausible
explanation could be the presence of a cluster of massive stellar
remnants. By applying equation.~\ref{eq:6}, we find that an equilibrium
configuration is possible under certain conditions.

In this study relativistic effects were not included for the orbit of
the star S2. The peri-centre-shift induced by relativistic effects is
of the order of 9~arcmin compared to the 40~arcmin expected from the
Newtonian shift due to the presence of an extended mass (see also
Rubilar \& Eckart, 2001, Fragile, \& Mathews, 2000).  It should be
pointed out that the peri-centre-shift due to relativistic effects is
in the opposite direction of that due to the extended mass and that
relativistic effects could in a way mask the amount of extended mask
around the black hole.  Therefore it will be important for future work
to investigate relativistic effects in more detail.

After only a few additional observational 
epochs it will be possible to put even stronger constraints on the 
mass of the cusp, and consequently to know more about the mass 
carrying population in the immediate vicinity of the black hole 
at the centre of the Milky Way.

As discussed in Rubilar \& Eckart (2001) a complete determination of the 
cusp parameters can be expected from combinations of 3 or more orbits
since then the number of degrees of freedom for the fits is increasing 
rapidly.
Further studies with more general density distributions should be
undertaken, because they may influence the precise value of the
extended mass contribution.
\begin{appendix}
\section{Orbital parameters of three 1~$\sigma$ fitting orbits}
In table~A.1, we give the orbital parameters of 3 exemplary orbits  
       giving fits within a 1~$\sigma$ limit.
      The orbits corresponding to those fits are plotted in Fig.6.   
      These are chosen for the case where the central black hole is present 
at the \emph{offset} position ((R.A~=~0.09~mpc; Dec~=~-0.12~mpc) from the 
central nominal radio position of Sgr~A*). 
      We also give the full range of parameters [Min;Max] which gives fits
corresponding to a 5$\sigma$ level.
             Here, we exclude solutions with M$_{tot}$~$>$~5.4$\times$10$^6$~M$_{\odot}$.
  
    \begin{table*} 
      \centering  
     \caption{Orbital parameters of three exemplary 1~$\sigma$ fitting orbits.}
       \begin{tabular}[]{||l||l||l||l||l||l||}
            \hhline{|t:=:t:=:t:=:t:=:t:==:t|}  
            & & & &  Min & Max\\
          \hhline{||~||~||~||~|:==:|}
       Total Mass $M_{tot}$ (10$^6$~M$_{\odot}$)&3.65&4.10&4.80&3.28&5.40\\
       \hhline{|t:=:t:=:t:=:t:=:t:==:t|}  
       Point Mass $M_{BH}$ (10$^6$~M$_{\odot}$)&3.65&3.69&3.65&3.28&3.97\\
       \hhline{|t:=:t:=:t:=:t:=:t:==:t|}  
      Cusp Percentage($\%$)&0.0&10&25&0&35\\
       \hhline{|t:=:t:=:t:=:t:=:t:==:t|}  
      M$^{apo}$ : Cusp mass inside S2's orbit(10$^6$~M$_{\odot}$)& 0.001& 0.049& 0.133 & 0 & 0.26 \\
       \hhline{|t:=:t:=:t:=:t:=:t:==:t|} 
      $e$ : Eccentricity&0.88&0.882&0.881&0.868&0.897\\
       \hhline{|t:=:t:=:t:=:t:=:t:==:t|}  
       $i$ : Inclination($^{\circ}$)&-47.89&-47.88&-48.12&-46.86&-50.19\\
        \hhline{|t:=:t:=:t:=:t:=:t:==:t|}  
      $a$ : Semi-major axis(mpc)&4.65&4.66&4.59&4.52&4.79\\
        \hhline{|t:=:t:=:t:=:t:=:t:==:t|}  
      $T$ : Time of peri-centre passage&2002.327&2002.328&2002.333&2002.327&2002.337\\
        \hhline{|t:=:t:=:t:=:t:=:t:==:t|}  
       $\Omega$ : Angle of line of nodes($^{\circ}$)&45.3&45.38&44.59&42.17&46.1\\
        \hhline{|t:=:t:=:t:=:t:=:t:==:t|}
       $\omega$ : Angle to peri-centre($^{\circ}$)&145.1&144.84&145.6&144.1&147.7\\
         \hhline{|t:=:t:=:t:=:t:=:t:==:t|}  
        $P$ : Period (Years)&15.55&15.43&15.06&14.61&16.67\\
        \hhline{|t:=:t:=:t:=:t:=:t:==:t|}  
        Peri-Centre distance&0.557&0.552&0.55&0.511&0.6\\
         \hhline{|t:=:t:=:t:=:t:=:t:==:t|}  
       $\Delta$$\alpha$ : Peri-centre-shift per revolution&00$^{\circ}$00'00''&00$^{\circ}$40'48''& 01$^{\circ}$52'43''&00$^{\circ}$00'00''&03$^{\circ}$07'12''\\
           \hline  
           \hline
         \end{tabular}
        \label{Tab}
         \end{table*}       

\end{appendix}

\acknowledgements
N. Mouawad thanks M. Freitag for substantial discussions.
This work has been partly supported by the 
Deutsche Forschungsgemeinschaft (DFG)
via grant SFB439 at the University of Heidelberg
and SFB494 at the University of Cologne.

\end{document}